\RequirePackage{ifpdf}
\ifpdf % We are running pdfTeX in pdf mode
\documentclass[pdftex]{sigma}
\else
\documentclass{sigma}
\fi

\begin{document}

\renewcommand{\PaperNumber}{008}

\FirstPageHeading

\ShortArticleName{Spectra of Observables in the $q$-Oscillator and
$q$-Analogue of the Fourier Transform}

\ArticleName{Spectra of Observables in the $\boldsymbol{q}$-Oscillator\\ and
$\boldsymbol{q}$-Analogue of the Fourier Transform}

\Author{Anatoliy U. KLIMYK}

\AuthorNameForHeading{A.U. Klimyk}

\Address{Bogolyubov Institute for Theoretical Physics,
 14-b Metrologichna Str.,  03143 Kyiv, Ukraine}

\Email{\href{mailto:aklimyk@bitp.kiev.ua}{aklimyk@bitp.kiev.ua}} 

\ArticleDates{Received August 26, 2005, in final form October 19, 2005; Published online October 21, 2005}

\Abstract{Spectra of the position and momentum operators of the
Biedenharn--Macfarlane $q$-oscillator (with the main relation
$aa^+-qa^+a=1$) are studied when $q>1$. These operators are
symmetric but not self-adjoint. They have a one-parameter family
of self-adjoint extensions. These extensions are derived
explicitly. Their spectra and eigenfunctions are given. Spectra of
different extensions do not intersect. The results show that the
creation and annihilation operators $a^+$ and $a$ of the
$q$-oscillator for $q>1$ cannot determine a physical system
without further more precise definition. In order to determine a
physical system we have to choose appropriate self-adjoint
extensions of the position and momentum operators.}


\Classification{47B15; 81Q10; 81S05}

\section{Introduction}

Approximately 20 years ago the first papers on quantum groups and
quantum algebras have appeared. Quantum groups and quantum
algebras can be considered as $q$-deformations of semisimple Lie
groups and Lie algebras. Soon after that an appropriate
$q$-deformation of the quantum harmonic oscillator, related to the
quantum group $SU_q(2)$, was introduced (see \cite{[1]} and
\cite{[2]}). There exist several variants of the $q$-oscillator.
They are obtained from each other by some transformation (see
\cite{[3]} and \cite[Chapter 5]{[3a]}). As their relation to the
nilpotent part of the quantum algebra $U_q({\rm sl}_3)$, see in
\cite{[3b]}.

One of the main problems for different forms of the $q$-oscillator
is form of spectra of the main operators, such as the
Hamiltonian, the position operator, the momentum operator, etc.
There is no problem with a spectrum of the Hamiltonian $H=\frac 12
(aa^++a^+a)$: this spectrum is discrete and the corresponding
eigenvectors are easily determined. But in some cases (similar to
the case of representations of noncompact quantum algebras (see,
for example, \cite{[4]}) there are difficulties with spectra of
the position and momentum operators (see \cite{[5],[6],[7]}).
It was shown that if the position operator $Q=a^++a$ (or the
momentum operator $P={\rm i}(a^+-a)$) is not bounded, then this
symmetric operator is not essentially self-adjoint. Moreover, in
this case it has deficiency indices (1,\,1), that is, it has a
one-parameter family of self-adjoint extensions. Finding
self-adjoint extensions of a closed symmetric (but not
self-adjoint) operator is a complicated problem. We need to know
self-adjoint extensions in order to be able to find their spectra.
As we shall see, different self-adjoint extensions of $Q$ and $P$
have different spectra.

In this paper we study self-adjoint extensions of the position and
momentum operators $Q$ and $P$ for the $q$-oscillator with the
main relation
 \[
aa^+-qa^+a=1
 \]
when $q>1$. For these values of $q$, the operators $Q$ and $P$ are
unbounded and not essentially self-adjoint (for $q<1$, these
operators are bounded and, therefore, self-adjoint; they are
studied in \cite{[6]}). These operators can be represented in an
appropriate basis by a Jacobi matrix. This means that they can be
studied by means of properties of $q$-orthogonal polynomials
associated with them (see Section 2 below). These $q$-orthogonal
polynomials are expressed in terms of the $q^{-1}$-continuous
Hermite polynomials $h_n(x|q)$ introduced by R. Askey \cite{[8]}.
These polynomials correspond to an indeterminate moment problem
and, therefore, are orthogonal with respect to infinitely many
positive measures (see Section 2 below). Using orthogonality
measures for these polynomials we shall find spectra of
self-adjoint extensions of $Q$ and $P$. This paper is an extended
exposition of the results of the paper \cite{[8a]}.

It follows from conclusions of this paper that the creation and
annihilation operators $a^+$ and $a$ of the $q$-oscillator for
$q>1$ do not determine uniquely a physical system. In order to fix
a physical system we have to choose appropriate self-adjoint
extensions of the position and momentum operators. This fact must
be taken into account with respect to applications of $q$-oscillators with
$q>1$. Thus, we cannot operate with the creation and annihilation
operators of the $q$-oscillator so freely as in the case of the
usual quantum harmonic oscillator.

Below we use (without additional explanation) notations of the
theory of $q$-special functions (see \cite{[9]}). In order to
study the position and momentum operators $Q$ and $P$ we shall
need the results on Jacobi matrices, orthogonal polynomials and
symmetric operators, representable by a Jacobi matrix. In the next
section, we give a combined exposition of some results on this
connection from books \cite[Chapter VII]{[10]}, \cite{[11]} and
from paper \cite{[12]} in the form, appropriate for a use below,
and some consequences of them.

\section{Operators representable by a Jacobi matrices,\\
and orthogonal polynomials}

\subsection{Jacobi matrices and orthogonal polynomials}

Many operators used in theoretical and mathematical physics are
operators representable by a~Jacobi matrix. There exists a
well-developed mathematical method for studying such operators.

In what follows we shall use only symmetric Jacobi matrices and
the word ``symmetric'' will be often omitted. By a symmetric Jacobi
matrix we mean a (finite or infinite) symmetric matrix of the form
 \begin{gather}
M =\left( \begin{matrix}
 b_{0}&  a_{0} &0     &0&0&  \cdots   \\
 a_{0}&  b_1   &a_{1} &0&0&  \cdots   \\
 0   &   a_1   &b_{2} &a_2&0&  \cdots  \\
 0   &   0    &a_{2} &b_3&a_3&  \cdots  \\
 \cdots&\cdots &\cdots &\cdots &\cdots &\cdots  \end{matrix}
 \right) . \label{eq2.1}
 \end{gather}
We assume below that $a_i\ne 0$, $i=0,1,2,\ldots$. All $a_i$ are
real. Let $L$ be a closed symmetric operator on a Hilbert space
${\cal H}$, representable by a Jacobi matrix $M$. Then there
exists an orthonormal basis $e_n$, $n=0,1,2,\ldots$, in ${\cal
H}$ such that
 \begin{gather}
 Le_n=a_ne_{n+1}+b_ne_n+a_{n-1}e_{n-1}, \label{eq2.1a}
 \end{gather}
where $e_{-1}\equiv 0$. Let
 \[
|x\rangle =\sum_{n=0}^\infty p_n(x)e_n
 \]
be an eigenvector\footnote{See that eigenvectors of $L$ may
belong to either the Hilbert space ${\cal H}$ or to some extension
of ${\cal H}$. (For example, if ${\cal H}=L^2(-\infty, \infty)$
and instead of $L$ we have the operator $d/dx$, then the functions
$e^{{\rm i}xp}$, which do not belong to $L^2(-\infty, \infty)$,
are eigenfunctions of $d/dx$.) Below we act freely with
eigenvectors, which do not belong to ${\cal H}$. The corresponding
reasoning can be easily made mathematically strict.} of $L$ with
an eigenvalue $x$, that is, $L| x\rangle =x|x\rangle$. Then
 \begin{gather*}
 L|x\rangle =\sum_{n=0}^\infty [  p_n(x) a_n e_{n+1}+p_n(x)
b_n e_{n}+p_n(x) a_{n-1} e_{n-1}]   \\
\phantom{L|x\rangle}{}=x \sum_{n=0}^\infty p_n(x) e_n.
 \end{gather*}
Equating coefficients at the vector $e_n$ one comes to a
recurrence relation for the coefficients~$p_n(x)$:
  \begin{gather}
a_np_{n+1}(x)+b_np_{n}(x)+a_{n-1}p_{n-1}(x)=xp_n(x). \label{eq2.2}
  \end{gather}
Since $p_{-1}(x)=0$, by setting $p_0(x)\equiv 1$ we see that
$p_1(x)=a_0x-b_0/a_0$. Similarly we can find successively
$p_n(x)$, $n=2,3,\ldots$. Thus, the relation \eqref{eq2.2}
completely determines the coefficients~$p_n(x)$. Moreover, the
recursive computation of $p_n(x)$ shows that these coefficients
$p_n(x)$ are polynomials in $x$ of degrees $n$, respectively.
Since the coefficients $a_n$ and $b_n$ are real (as the
matrix $M$ is symmetric),  all coefficients of the polynomials
$p_n(x)$ themselves are real.

Since $a_n>0$ and $b_n\in {\mathbb R}$ in \eqref{eq2.2} then due to
Favard's characterization theorem the polyno\-mials~$p_n(x)$ are
ortho\-gonal with respect to some positive measure $\mu(x)$. It is
known that ortho\-gonal polynomials admit orthogonality with respect
to either unique positive measure or with respect to infinitely
many positive measures.

The polynomials $p_n(x)$ are very important for studying
properties of the closed symmetric operator $L$. Namely, the
following statements are true (see, for example, \cite{[10]} and
\cite{[12]}):
\medskip

I. Let the polynomials $p_n(x)$ be orthogonal with respect to a
unique orthogonality measure~$\mu$,
 \[
\int_{\mathbb R}  p_m(x)p_n(x)d\mu(x)=\delta_{mn}.
 \]
Then the corresponding closed operator $L$ is self-adjoint.
Moreover, the spectrum of the operator~$L$ is simple and coincides
with the set, on which the polynomials $p_n(x)$ are orthogonal
(that is, with the support of the measure $\mu$). The measure
$\mu(x)$ determines also the spectral measure for the operator $L$
(for details see \cite[Chapter VII]{[10]}).

 \medskip

II. Let the polynomials $p_n(x)$ be orthogonal with respect to
infinitely many different orthogo\-nality measures $\mu$. Then the
closed symmetric operator $L$ is not self-adjoint and has
deficiency indices (1,\,1), that is, it has infinitely many (in
fact, one-parameter family of) self-adjoint extensions. It is
known that among orthogonality measures, with respect to which the
polynomials are orthogonal, there are so-called {\it extremal}
measures (that is, such measures that a set of polynomials $\{
p_n(x)\}$ is complete in the Hilbert space $L^2$ with respect to
the corresponding measure; see Subsection 2.3 below). These
measures uniquely determine self-adjoint extensions of the
symmetric operator $L$. There exists one-to-one correspondence
between essentially distinct extremal orthogonality measures and
self-adjoint extensions of the operator $L$. The extremal
orthogonality measures determine spectra of the corresponding
self-adjoint extensions.
 \medskip

The inverse statements are also true:
 \medskip

I$'$. Let the operator $L$ be self-adjoint. Then the corresponding
polynomials $p_n(x)$ are ortho\-gonal with respect to a unique
orthogonality measure $\mu$,
 \[
\int_{\mathbb R} p_m(x)p_n(x)d\mu(x)=\delta_{mn},
 \]
where a support of $\mu$ coincides with the spectrum of $L$.
Moreover, a measure $\mu$ is uniquely determined by a spectral
measure for the operator $L$ (for details see \cite[Chapter VII]{[10]}).

 \medskip

II$'$. Let the closed symmetric operator $L$ be not self-adjoint.
Since it is representable by a~Jacobi matrix \eqref{eq2.1} with
$a_n\ne 0$, $n=0,1,2,\ldots$, it admits a one-parameter family of
self-adjoint extensions (see \cite[Chapter VII]{[10]}). Then the
polynomials $p_n(x)$ are orthogonal with respect to infinitely
many orthogonality measures $\mu$. Moreover, spectral measures of
self-adjoint extensions of $L$ determine extremal orthogonality
measures for the polynomials $\{ p_n(x)\}$ (and a set of
polynomials $\{ p_n(x)\}$ is complete in the Hilbert spaces
$L^2(\mu)$ with respect to the corresponding extremal measures
$\mu$).
 \medskip

On the other hand, with the orthogonal polynomials $p_n(x)$,
$n=0,1,2,\ldots$ the classical moment problem is associated (see
\cite{[11]} and \cite{[13]}). Namely, with these polynomials (that
is, with the coefficients $a_n$ and $b_n$ in the corresponding
recurrence relation) real numbers $c_n$, $n=0,1,2,\ldots$ are
associated, which determine the corresponding classical moment
problem. (The numbers $c_n$ are uniquely determined by $a_n$ and
$b_n$, see \cite{[11]}.)

The definition of the classical moment problem consists in the
following. Let a set of real numbers $c_n$, $n=0,1,2,\ldots$ be
given. We are looking for a positive measure $\mu(x)$, such that
  \begin{gather}
\int x^nd\mu(x)=c_n,\qquad n=0,1,2,\ldots , \label{eq2.3}
  \end{gather}
where the integration is taken over ${\mathbb R}$ (in this case we
deal with the {\it Hamburger moment problem}). There are two
principal questions in the theory of moment problem:

 \medskip

(i) Does there exist a measure $\mu(x)$, such that relations
\eqref{eq2.3} are satisfied?

(ii) If such a measure exists, is it determined uniquely?

 \medskip

The answer to the first question is positive, if the numbers
$c_n$, $n=0,1,2,\ldots$ are those corresponding to a family
of orthogonal polynomials. Moreover, then the measure $\mu(x)$
coincides with the measure with respect to which these
polynomials are orthogonal.

If a measure $\mu$ in \eqref{eq2.3} is determined uniquely, we say
that we deal with {\it the determinate moment problem}. (In
particular, it is the case when the measure $\mu$ is supported on
a bounded set.) Then the corresponding polynomials $\{ p_n(x)\}$
are orthogonal with respect to this measure and {\it the
corresponding symmetric operator $L$ is self-adjoint}.

 If a measure with respect to which relations
\eqref{eq2.3} hold is not unique, then we say that we deal with
{\it the indeterminate moment problem}. In this case there exist
infinitely many measures~$\mu(x)$ for which \eqref{eq2.3} take
place. Then the corresponding polynomials are orthogonal with
respect to all these measures, and {\it the corresponding symmetric
operator $L$ is not self-adjoint}. In this case the set of
solutions of the moment problem for the numbers $\{ c_n\}$
coincides with the set of orthogonality measures for the
corresponding polynomials $\{ p_n(x)\}$.

See that not every set of real numbers $c_n$, $n=0,1,2,\ldots$
is associated with a set of orthogonal polynomials. In other
words, there are sets of real numbers $c_n$, $n=0,1,2,\ldots$
such that the corresponding moment problem does not have a
solution, that is, there is no positive measure~$\mu$, for which
the relations \eqref{eq2.3} are true. But if for some set of real
numbers $c_n$, $n=0,1,2,\ldots$ the moment problem \eqref{eq2.3}
has a solution $\mu$, then this set corresponds to some set of
polyno\-mials~$p_n(x)$, $n=0,1,2,\ldots$, which are orthogonal with
respect to this measure $\mu$. There exist criteria indicating
when for a given set of real numbers $c_n$, $n=0,1,2,\ldots$ the
moment problem~\eqref{eq2.3} has a~solution (see, for example,
\cite{[11]}). Moreover, there exist procedures, that associate a
collection of orthogonal polynomials to a set of real numbers
$c_n$, $n=0,1,2,\ldots$ for which the moment problem
\eqref{eq2.3} has a solution (see, \cite{[11]}).

\medskip

Thus, we see that the following three theories are closely
related:
\medskip

(i) the theory of symmetric operators $L$, representable by a
Jacobi matrix;

(ii) the theory of orthogonal polynomials in one variable;

(iii) the theory of classical moment problem.

\subsection{Self-adjointness}

We have seen that orthogonal polynomials $\{p_n(x)\}$ associated
with the symmetric operator~$L$ determine whether it is
self-adjoint or not. There exist other criteria of
self-adjointness of a~closed symmetric operator:

 \medskip

(a) If the coefficients $a_n$ and $b_n$ in \eqref{eq2.1a} are
bounded, the operator $L$ is bounded and, therefore, self-adjoint.
 \medskip

(b) If in \eqref{eq2.1a} $b_n$ are arbitrary real numbers and $a_n$ are
such that
 \[
\sum_{n=0}^\infty a_n^{-1}=\infty ,
 \]
then the operator $L$ is self-adjoint.
 \medskip

(c) Let $|b_n|\le C$, $n=0,1,2,\ldots$, and let for some positive
$j$ we have $a_{n-1}a_{n+1}\le a_n^2$, $n\ge j$. If
\[
\sum_{n=0}^\infty a_n^{-1}<\infty ,
 \]
then the operator $L$ is not self-adjoint.
 \medskip

Each closed symmetric operator representable by a Jacobi matrix,
which is not a self-adjoint operator, has deficiency indices (1,\,1), 
that is, it has a one-parameter family of self-adjoint
extensions. Let us explain this statement in more detail. Let $A$
be a closed operator (not obligatorily symmetric) on a Hilbert space
${\cal H}$. Suppose that a domain $D(A)$ of $A$ is an everywhere
dense subspace of ${\cal H}$. There are pairs $v$ and $v'$ of
elements from ${\cal H}$ such that
 \[
\langle v| A u\rangle =\langle v'|u\rangle \quad {\rm for\ all}
\quad u\in D(A).
 \]
We set $v'=A^*v$ and call $A^*$ the operator conjugate to $A$. It
is proved that $A^*$ is defined on an everywhere dense subspace
$D(A^*)$ of ${\cal H}$ (since $D(A)$ is everywhere dense in ${\cal
H}$). The operator $A^*$ is linear and closed. Moreover,
${A^*}^*=A$. It is called {\it symmetric} if $A\subseteq A^*$,
that is, $D(A)\subseteq D(A^*)$ and $A$ coincides with $A^*$ on
$D(A)$. If $A=A^*$, that is, $D(A)=D(A^*)$, then $A$ is called
{\it self-adjoint}. If a closed symmetric operator $A$ is not
self-adjoint, that is, $D(A)\subset D(A^*)$ and $D(A)\ne D(A^*)$,
then it can have self-adjoint extensions. The fact that {\it the
operator $A$ has a self-adjoint extension $A^{\rm ext}$ means that
$(A^{\rm ext})^*=A^{\rm ext}$ (that is, $A^{\rm ext}$ is a
self-adjoint operator), $D(A)\subset D(A^{\rm ext})$ and the
operators $A$ and $A^{\rm ext}$ coincide on $D(A)$.}

 To know whether a symmetric operator $A$ has self-adjoint extensions
or not, there is the notion of deficiency indices $(m,n)$ of $A$
($m$ and $n$ are nonnegative integers). If these indices are equal
to each other, then $A$ has self-adjoint extensions. Self-adjoint
extensions are constructed with the help of two deficiency
subspaces (they are of dimensions $m$ and $n$, respectively). 
A~detailed description of deficiency indices and deficiency
subspaces can be found in \cite{[15]}.

The important fact is that different self-adjoint extensions of a
symmetric operator can have different spectra (this can happen
when their spectra have discrete parts). We shall meet this
situation in the forthcoming sections.

\subsection{Extremal orthogonality measures}

As we have seen in Subsection 2.1, a closed symmetric operator $L$
representable by a Jacobi matrix that is not self-adjoint can
be studied by means of orthogonal polynomials, associated with
$L$. Self-adjoint extensions of $L$ are connected with extremal
orthogonality measures for these polynomials. Let us consider such
measures in more detail.

Let $p_n(x)$, $n=0,1,2,\ldots$ be a set of orthogonal polynomials
associated with an indeterminate moment problem \eqref{eq2.3}.
Then for orthogonality measures $\mu (x)$ for these polynomials
the following formula holds:
  \begin{gather}
F(z)\equiv \frac{A(z)-\sigma(z)C(z)}{B(z)-\sigma(z)D(z)}
=\int_{-\infty}^\infty \frac{d\mu(t)}{z-t}  \label{eq2.4}
  \end{gather}
where $A(z), B(z), C(z), D(z)$ are entire functions which are the
same for all orthogonality measures $\mu$. These functions are
related to asymptotics of the polynomials $p_n(x)$ and of an
associated set of polynomials $p^*_n(x)$ (see \cite{[11]} for details).
Expressions for $A(z)$, $B(z)$, $C(z)$, $D(z)$ as infinite sums (in $n$)
of the polynomials $p_n(x)$ and $p^*_n(x)$ see \cite[section
VII.7]{[10]}. In \eqref{eq2.4}, $\sigma(z)$ is a~Nevanlinna function. Moreover, to
each such function $\sigma(z)$ (including cases of constant
$\sigma(z)$ and $\sigma(z)=\pm \infty$) there is a corresponding single
orthogonality measure $\mu(t)\equiv \mu_\sigma (t)$ and,
conversely, to each orthogonality measure $\mu$ there is a corresponding
function $\sigma$ such that formula~\eqref{eq2.4} holds. There
exists the Stieltjes inversion formula that converts the formula
\eqref{eq2.4}. It has the form
 \begin{gather*}
[\mu(t_1+0)+\mu(t_1-0)]-[\mu(t_0+0)+\mu(t_0-0)] \\
\qquad {}= \lim_{\varepsilon\to +0}\left( -\frac{1}{\pi{\rm i}}
\int_{t_0}^{t_1} [F(t+{\rm i}\varepsilon)-F(t-{\rm
i}\varepsilon)]dt \right) .
\end{gather*}

Thus, orthogonality measures for a given set of polynomials
$p_n(x)$, $n=0,1,2,\ldots$ in principle can be found. However,
it is very difficult to evaluate the functions $A(z)$, $B(z)$,
$C(z)$, $D(z)$. (In~\cite{[14]} they are evaluated for particular
example of polynomials, namely, for the $q^{-1}$-continuous
Hermite polynomials $h_n(x|q)$.) So, as a rule, for the derivation
of orthogonality measures other methods are used.

The measures $\mu_\sigma (t)$, corresponding to constants $\sigma$
(including $\sigma=\pm \infty$), are called {\it extremal
measures} (some authors, following the book \cite{[13]}, call
these measures $N$-extremal). All other orthogonality measures are
not extremal.

The importance of extremal measures is explained by Riesz's
theorem. Let us suppose that a set of polynomials $p_n(x)$,
$n=0,1,2,\ldots$ associated with the indeterminate moment
problem, is orthogonal with respect to a positive measure $\mu$
(that is, $\mu$ is a solution of the moment 
problem~\eqref{eq2.3}). Let $L^2(\mu)$ be the Hilbert space of square
integrable functions with respect to the measure $\mu$. Evidently,
the polynomials $p_n(x)$ belong to the space $L^2(\mu)$. Riesz's
theorem states the following:

  \medskip

\noindent
{\bf Riesz's theorem.} {\it The set of polynomials $p_n(x)$,
$n=0,1,2,\ldots$ is complete in the Hilbert space $L^2(\mu)$
(that is, they form a basis in this Hilbert space) if and only if
the measure $\mu$ is extremal.}
 \medskip

Note that if a set of polynomials $p_n(x)$, $n=0,1,2,\ldots$
corresponds to a determinate moment problem and $\mu$ is an
orthogonality measure for them, then this set of polynomials is
also complete in the Hilbert space $L^2(\mu)$.

 Riesz's theorem is often used in order to determine whether a
certain orthogonality measure is extremal or not. Namely, if we
know that a given set of orthogonal polynomials corresponding to
an indeterminate moment problem is not complete in the Hilbert
space $L^2(\mu)$, where $\mu$ is an orthogonality measure, then
this measure is not extremal.

Note that for applications in physics and in functional analysis
it is of interest to have extremal orthogonality measures. If an
orthogonality measure $\mu$ is not extremal, then it is important
to find a system of orthogonal functions $\{f_m(x)\}$, which
together with a given set of polynomials constitute a complete set
of orthogonal functions (that is, a basis in the Hilbert space
$L^2(\mu)$). Sometimes, it is possible to find such systems of
functions (see, for example, \cite{[16]}).

Extremal orthogonality measures have many interesting 
properties~\cite{[11]}:

 \medskip

(a) An extremal measure $\mu_\sigma(x)$ associated (according to
formula~\eqref{eq2.4}) with a number $\sigma$ is discrete. Its
spectrum (that is, the set on which the corresponding polynomials
$p_n(x)$, $n=0,1,2,\ldots$ are orthogonal) coincides with the set
of zeros of the denominator $B(z)-\sigma D(z)$ in~\eqref{eq2.4}.
The mass concentrated at a spectral point $x_j$ (that is, a jump
of $\mu_\sigma(x)$ at the point $x_j$) is equal to
$\big(\sum\limits_{n=0}^\infty | p_n(x_j)|^2\big)^{-1}$.

\medskip

(b) Spectra of extremal measures are real and simple. This means
that the corresponding self-adjoint operators, which are
self-adjoint extensions of the operator $L$, have simple spectra,
that is, all spectral points are of multiplicity 1.
\medskip

(c) Spectral points of two different extremal measures
$\mu_\sigma(x)$ and $\mu_{\sigma'}(x)$ are mutually separated.
\medskip

(d) For a given real number $x_0$, there always exists a (unique) real
value $\sigma_0$, such that the measure $\mu_{\sigma_0}(x)$ has
$x_0$ as its spectral point. The points of the spectrum of
$\mu_\sigma(x)$ are analytic monotone functions of $\sigma$.
\medskip

It is difficult to find all extremal orthogonality measures for a
given set of orthogonal polynomials (that is, self-adjoint
extensions of a corresponding closed symmetric operator). As far
as we know, at present they are known only for one family
of polynomials, which correspond to indeterminate moment problem.
They are the $q^{-1}$-continuous Hermite polynomials $h_n(x|q)$
(see \cite{[14]}).

As noted in \cite{[11]}, if extremal measures $\mu_\sigma$ are known
then by multiplying $\mu_\sigma$ by a suitable factor (depending
on $\sigma$) and integrating it with respect to $\sigma$, one can
obtain infinitely many continuous orthogonality measures (which
are not extremal).

Since spectra of self-adjoint extensions of the operator $L$
coincide with  the spectra of the corresponding extremal
orthogonal measures for the polynomials $p_n(x)$, then the
properties (a)--(d) can be formulated for spectra of these
self-adjoint extensions:
 \medskip

(a$'$) Spectra of self-adjoint extensions of $L$ are discrete.

(b$'$) Self-adjoint extensions of $L$ have simple spectra, that
is, spectral points are not multiple.

(c$'$) Spectra of two different self-adjoint extensions of $L$ are
mutually separated.

(d$'$) For a given real number $x_0$, there exists a (unique)
self-adjoint extension $L^{\rm ext}$ such that $x_0$ is a spectral
point of $L^{\rm ext}$.

\section[The Biedenharn-Macfarlane $q$-oscillator]{The Biedenharn--Macfarlane ${\boldsymbol q}$-oscillator}

There are different forms of the $q$-oscillator algebra (see
\cite{[3]}; some forms of $q$-oscillators go back to the paper by
R. Santilli \cite{[17]}) which mathematically are not completely
equivalent. For our definition of the $q$-oscillator we use the
following relations
 \[
aa^+ -qa^+a=1,\qquad [N,a^+]=a^+,\qquad [N,a]=-a
 \]
for the creation and annihilation operators $a^+$, $a$ and for the
number operator $N$.

The Fock representation of this $q$-oscillator acts on the Hilbert
space ${\cal H}$ with the orthonormal basis $|n\rangle$,
$n=0,1,2,\ldots$, and is given by the formulas
 \begin{gather}
 a|n\rangle =
\{n\}_{q}^{1/2}|n - 1\rangle,\qquad a^+|n\rangle =
\{n+1\}_{q}^{1/2}|n+1\rangle,\qquad N |n\rangle =n|n\rangle ,
\label{eq5}
 \end{gather}
where the expression
  \begin{gather}
 \{n\}_{q} := \frac{q^n-1}{q-1}  \label{eq5a}
  \end{gather}
is called a $q$-{\it number}.

Note that the basis vectors $|n\rangle$ are eigenvectors of the
Hamiltonian $H=\frac 12 (aa^++a^+a)$. It follows from \eqref{eq5}
that
 \[
 H|n\rangle =\frac 12 (\{ n+1\}_q + \{ n\}_q) |n\rangle=\frac 12
 \frac{q^n(q+1)-2}{q-1} |n\rangle .
  \]
Thus, the spectrum of $H$ consists of the points $\frac 12
 (q^n(q+1)-2)/(q-1)$, $n=0,1,2,\ldots$.

\section{Functional realization of the Fock representation}

The Fock representation of the $q$-oscillator can be realized on
many spaces of functions. We shall need the realization related to
the space of polynomials in one variable.

Let ${\cal P}$ be the space of all polynomials in a variable $y$
(this variable has no relation to the coordinates of the position
operator $Q$ considered below). We introduce in ${\cal P}$ a
scalar product such that the monomials
 \begin{gather}
 e_n\equiv e_n(y):=
\frac{(-1)^{n/2}}{(q;q)_n^{1/2}} y^n,   \label{eq6}
 \end{gather}
where
 \[
 (b;q)_n:=(1-b)(1-bq)\cdots (1-bq^{n-1}),
 \]
constitute an orthonormal basis of ${\cal P}$. The orthonormality
of this basis gives a scalar product in~${\cal P}$. We close the
space ${\cal P}$ with respect to this scalar product and obtain a
Hilbert space which can be considered as a realization of the
Hilbert space ${\cal H}$. The operators $a^+$ and $a$ are realized
on this space as
 \[
a^+=(q-1)^{-1/2} y,\qquad a=(q-1)^{1/2} D_q,
 \]
where $D_q$ is the {\it $q$-derivative} determined by
 \[
D_qf(y)=\frac{f(qy)-f(y)}{(q-1)y}.
 \]
Then the operators $a^+$ and $a$ act upon the basis elements
\eqref{eq6} by formulas \eqref{eq5}. Everywhere below we assume
that ${\cal H}$ is the Hilbert space of functions in $y$,
introduced above.

\section{Position and momentum operators}

We are interested in the position and momentum operators
 \[
Q=a^++a, \qquad P={\rm i}(a^+-a)
 \]
of the $q$-oscillator. It is clear from \eqref{eq5a} that
  \begin{gather}
 Q=(q-1)^{-1/2} y+(q-1)^{1/2} D_q,  \label{eq6b}\\
   P={\rm i}(q-1)^{-1/2} y-{\rm i}(q-1)^{1/2} D_q.\nonumber
 \end{gather}
We have the formulas
 \begin{gather}
Qe_n=\{n\}_q^{1/2}e_{n-1}+\{n+1\}_q^{1/2}e_{n+1},  \label{eq7}
\\
Pe_n={\rm i}\{n\}_q^{1/2}e_{n-1}-{\rm i}\{n+1\}_q^{1/2}e_{n+1},
 \label{eq8}
  \end{gather}
which follow from \eqref{eq5}. When $q<1$, it is clear from
the definition \eqref{eq5a} of $\{ n\}_q$ that $Q$ and~$P$ are
bounded and therefore self-adjoint, operators. When $q>1$, then it
follows from \eqref{eq7} and~\eqref{eq8} that $Q$ and $P$ are
unbounded symmetric operators. Since $\{ n-1\}\{ n+1\}\le \{
n\}^2$, then it follows from criterion (c) of Subsection 2.2 that
closures of these operators are not self-adjoint. Each of these
operators has a one-parameter family of self-adjoint extensions.
One of the aims of this paper is to give these self-adjoint
extensions using reasoning of Section 2.

\section{Eigenfunctions of the position operator}

{\bf We assume below that $\boldsymbol{q}$ is a fixed real number such that
$\boldsymbol{q>1}$}. For convenience, we also introduce the notation $\breve q
=q^{-1}$.

The aim of this section is to derive formulas for eigenfunctions
$\varphi_x(y)$ of the position operator~$Q$:
 \[
Q \varphi_x(y)=x \varphi_x(y).
 \]
Let us show that
 \begin{gather}
 \varphi_x(y) =\prod_{n=0}^\infty \left(1+2y x'
{\breve q}^{n+1}-y^2{\breve q}^{2n+2}\right),   \label{eq9}
  \end{gather}
where $x':=\frac 12 (q-1)^{1/2}x$. Using formula \eqref{eq6b} and
the definition of the $q$-derivative $D_q$ we have
 \[
D_q \varphi_x(y) =\frac{\varphi_x(qy)-\varphi_x(y)}{y(q-1)}
=\frac{2x'-y}{q-1} \varphi_x(y).
 \]
Therefore,
 \[
 Q\varphi_x(y)=\left\{
\frac{y}{(q-1)^{1/2}}+\frac{2x'-y}{(q-1)^{1/2}} \right\}
\varphi_x(y)=\frac{2x'}{(q-1)^{1/2}}= x\varphi_x(y)
 \]
that is, the functions \eqref{eq9} are eigenfunctions of the
operator $Q$.

Let us reduce the functions \eqref{eq9} to another form. To do
this we note that
 \[
1+2yx'{\breve q}^{n+1}-y^2{\breve q}^{2n+2}=\left(1+{\breve q}^ny\breve
q \left(\sqrt{1+{x'}^2}+x'\right)\right) \left(1-{\breve q}^ny\breve q
\left(\sqrt{1+{x'}^2}-x'\right)\right).
 \]
Thus,
 \begin{gather*}
\varphi_x(y) = \prod_{n=0}^\infty \left(1+{\breve q}^ny\breve q
\left(\sqrt{1+{x'}^2}+x'\right)\right) \left(1-{\breve q}^ny\breve q
\left(\sqrt{1+{x'}^2}-x'\right)\right)  \\
\phantom{\varphi_x(y)}{} =\left(-y\breve q \left(\sqrt{1+{x'}^2}+x'\right);\breve q\right)_\infty \left(y\breve q
\left(\sqrt{1+{x'}^2}-x'\right);\breve q\right)_\infty .
 \end{gather*}
Comparing the right hand side of the above formula with the right hand side in the
formula
 \[
\sum_{n=0}^\infty \frac{t^n {\breve q}^{n(n-1)/2}}{({\breve
q};{\breve q})_n} h_n(y|{\breve q})=\left(-t\left(\sqrt{y^2+1} +y\right);{\breve
q}\right)_\infty \left(t\left(\sqrt{y^2+1} -y\right);{\breve q}\right)_\infty
 \]
(see formula (2.4) in \cite{[14]}), giving a generating function
for the $q^{-1}$-Hermite polynomials $h_n(x|\breve q)$ defined by
 \[
h_n(x|\breve q)=\sum _{k=0}^n \frac{(-1)^k\breve q^{k(k-n)}
(\breve q;\breve q)_n}{(\breve q;\breve q)_k(\breve q;\breve
q)_{n-k}}\left(\sqrt{x^2+1}+x\right)^{n-2k},
 \]
we conclude that the functions $\varphi_x(y)$ can be decomposed into
the orthogonal polynomials $h_n(x'|\breve q)$ (other expressions
for $h_n(x|\breve q)$ can be obtained from expressions for the
continuous $q$-Hermite polynomials $H_n(x|q)$ in \cite{[13a]}
since $H_n({\rm i}x|\breve q)={\rm i}^n h_n(x|\breve q)$, see
\cite{[8]}). We have
 \begin{gather}
\varphi_x(y)=\sum_{n=0}^\infty \frac{y^n{\breve
q}^{n(n+1)/2}}{(\breve q;\breve q)_n} h_n(x'|\breve q).
 \label{eq10}
 \end{gather}
Taking into account the expression \eqref{eq6} for the basis
elements $e_n$ and the formula
 \[
(\breve q;\breve q)_n=(-1)^n(q;q)_nq^{-n(n+1)/2}, \qquad \breve q
=q^{-1},
 \]
we derive that
 \begin{gather*}
\varphi_x(y) =\sum_{n=0}^\infty \frac{(-1)^ny^n}{(q;q)_n}
h_n(x'|\breve q) \\
\phantom{\varphi_x(y)}{}=\sum_{n=0}^\infty
\frac{(-1)^n(-1)^{-n/2}e_n(y)}{(q;q)_n^{1/2}}h_n(x'|\breve q) \\
\phantom{\varphi_x(y)}{}=\sum_{n=0}^\infty e_n(y) \frac{(-1)^n{\breve
q}^{n(n+1)/4}}{(\breve q;\breve q)_n^{1/2}}h_n(x'|\breve q).
 \end{gather*}
Thus, we proved the following decomposition of the eigenfunctions
$\varphi_x(y)$ in the basis elements~\eqref{eq6} of the Hilbert
space ${\cal H}$:
 \begin{gather}
\varphi_x(y)=\sum_{n=0}^\infty P_n(x)e_n(y),  \label{eq11}
 \end{gather}
where the coefficients $P_n(x)$ are given by the formula
 \begin{gather}
P_n(x)=(-1)^n{\breve q}^{n(n+1)/4}(\breve q;\breve
q)_n^{-1/2}h_n(x'|\breve q),  \label{eq12}
  \end{gather}
and, as before, $x'=\frac 12 (q-1)^{1/2}x$.

We have found that eigenfunctions of the position operator $Q$ are
given by formula \eqref{eq10}. However, we do not know the spectra
of self-adjoint extensions $Q^{\rm ext}$ of $Q$. In order to find
these extensions and their spectra we use assertions of Section 2.
Namely, since eigenfunctions of $Q$ are expressed in terms of the
basis elements $e_n(y)$ by formula \eqref{eq11}, then self-adjoint
extensions $Q^{\rm ext}$ and their spectra are determined by
orthogonality relations of the polynomials \eqref{eq12}.

\section[Spectra of self-adjoint extensions of $Q$]{Spectra of self-adjoint extensions of $\boldsymbol{Q}$}

The polynomials $h_n(z|\breve q)$, $n=0,1,2,\ldots$, $0<\breve
q<1$, have infinitely many orthogonality relations. Extremal
orthogonality measures are parametrized by a real number $b$,
$\breve q\le b<1$, which is related to the parameter $\sigma$ of
Section 2 (see \cite{[14]}). It is shown in \cite{[14]} that for a
fixed $b$, the corresponding orthogonality measure is concentrated
on the discrete set of points
 \[   
 z_b(r)=\frac 12 ({\breve
q}^{-r}b^{-1}-b{\breve q}^r),\qquad r=0,\pm 1,\pm 2, \ldots,
 \]
and the orthogonality relation is given by
 \begin{gather}
\sum_{r=-\infty} ^\infty m_rh_n(z_b(r)|\breve
q)h_{n'}(z_b(r)|\breve q)={\breve q}^{-n(n+1)/2}(\breve q;\breve
q)_n \delta_{nn'},  \label{eq14}
 \end{gather}
where the weight function $m_r$ coincides with
  \begin{gather}
m_r=\frac{b^{4r}{\breve q}^{r(2r-1)}(1+b^2{\breve
q}^{2r})}{(-b^2;\breve q)_\infty (-\breve q/b^2;\breve q)_\infty
(\breve q;\breve q)_\infty}  \label{eq15}
 \end{gather}
and
 \[
 (a;\breve q)_\infty :=\prod_{s=0}^\infty (1-a\breve q^s).
 \]

Therefore, the orthogonality relations for the polynomials
\eqref{eq12} with extremal orthogo\-na\-lity measures are given by the
same parameter $b$, $\breve q\le b<1$, and for fixed $b$ the
measure is concentrated on the discrete set
 \begin{gather}
x_b(r)=({\breve q}^{-r}b^{-1}-b{\breve q}^r)/(q-1)^{1/2},\qquad
r=0,\pm 1,\pm2,\ldots .   \label{eq16}
  \end{gather}
The corresponding orthogonality relation is
 \begin{gather}
\sum_{r=-\infty}^\infty m_rP_n(x_b(r))P_{n'}(x_b(r))=\delta_{nn'},
 \label{eq17}
 \end{gather}
where $m_r$ is given by \eqref{eq15}.

These orthogonality relations and assertions of Section 2 allow us
to make the following conclusions:

\begin{theorem}  Self-adjoint extensions $Q^{\rm ext}_b$ of the position
operator $Q$ are given by the parameter~$b$, $\breve q\le b<1$.
Moreover, the spectrum of the extension $Q^{\rm ext}_b$ coincides
with the set of points
 \begin{gather}
x_b(r)=(q^rb^{-1}-bq^{-r})/(q-1)^{1/2},\qquad r=0,\pm
1,\pm2,\ldots .   \label{eq18}
 \end{gather}
\end{theorem}

These points coincide with values of the coordinate of our
physical system fixed by the parameter $b$. It follows from
\eqref{eq12} that to the eigenvalues \eqref{eq18} there correspond
eigenfunctions
 \begin{gather}
\varphi_{x_b(r)}(y)=\sum_{n=0}^\infty P_n(x_r(b))e_n(y),\qquad
r=0,\pm 1,\pm 2,\ldots.   \label{eq19}
 \end{gather}

From \eqref{eq18} and from assertions of Section 2 we derive the
following corollary:
 \begin{corollary}
$({\rm a})$ Spectra of the self-adjoint extensions $Q^{\rm ext}_b$
are discrete and simple.

$({\rm b})$ Spectra of two different self-adjoint extensions
$Q^{\rm ext}_b$ and $Q^{\rm ext}_{b'}$, $b\ne b'$, are mutually
separated.

$({\rm c})$ For a given real number $x_0$ there exists a (unique)
self-adjoint extension $Q^{\rm ext}_b$ such that $x_0$ is a
spectral point of $Q^{\rm ext}_b$.
 \end{corollary}

For a fixed $b$, the eigenfunctions \eqref{eq19} are linearly
independent (and, therefore, orthogonal), since they correspond to
different eigenvalues of $Q^{\rm ext}_b$. Since the corresponding
orthogonality measure in \eqref{eq17} is extremal, they constitute
a basis of the Hilbert space ${\cal H}$. Let us normalize these
basis elements. To do this, we have to multiply each
$\varphi_{x_r(b)}(y)$ by the corresponding normalization constant:
 \[
\varphi^{\rm norm}_{x_b(r)}(y)  =c_b(r)\varphi_{x_b(r)}(y),\qquad
r=0,\pm 1,\pm 2,\ldots.
 \]
These functions form an orthonormal basis of ${\cal H}$. Since
 \begin{gather}
\varphi^{\rm norm}_{x_b(r)}(y)  =\sum_{n=0}^\infty
c_b(r)P_n(x_b(r))e_n(y)  \label{eqaa}
 \end{gather}
the matrix $(a_{rn})$, $a_{rn}=c_b(r)P_n(x_b(r))$, where $r=0,\pm
1,\pm 2,\ldots$ and $n=0,1,2,\ldots$ connects two orthonormal
bases of the Hilbert space ${\cal H}$. Therefore, this matrix is
unitary, that is,
 \[
\sum_{r=-\infty}^\infty |c_b(r)|^2
P_n(x_b(r))P_{n'}(x_b(r))=\delta_{nn'}.
 \]
Comparing this formula with relation \eqref{eq17} we have
$c_b(r)=m_r^{1/2}$ and
 \[
\varphi^{\rm norm}_{x_b(r)}(y)=m_r^{1/2} \varphi_{x_b(r)}(y),\qquad r=0,\pm 1,\pm 2,\ldots,
 \]
where $m_r\equiv m_r(b)$ is given by \eqref{eq15}.

\section[Coordinate realization of ${\cal H}$]{Coordinate realization of $\boldsymbol{{\cal H}}$}

In order to realize $Q^{\rm ext}_b$ as a self-adjoint operator, we
construct a one-to-one isometry $\Omega$ of the Hilbert space
${\cal H}$ onto the Hilbert space $L^2_b(m_r)$ of functions $F$ on
the set of points \eqref{eq18} (coinciding with the set of values
of the coordinate) with the scalar product
 \[
\langle F(x_b(r)),F'(x_b(r))\rangle =\sum_{r=-\infty}^\infty
m_rF(x_b(r))\overline{F'(x_b(r))}.
 \]
It follows from \eqref{eq17} that the polynomials $P_n(x_b(r))$
are orthogonal on the set \eqref{eq18} and constitute an
orthonormal basis of $L^2_b(m_r)$. For a fixed $b$, the isometry
$\Omega$ is given by the formula
 \[
\Omega: \ \ {\cal H}\ni f\to F(x_b(r)):=m_r^{-1/2}\langle
f,\varphi^{\rm norm}_{x_b(r)}(y) \rangle_{\cal H}\in L^2_b(m_r).
 \]
It follows from \eqref{eqaa} that
 \[
{\cal H}\ni e_n(y)\to m_r^{-1/2}\langle e_n(y),\varphi^{\rm
norm}_{x_b(r)}(y) \rangle_{\cal H} =P_n(x_b(r)).
 \]
This formula shows that $\Omega$ is indeed a one-to-one isometry.

The operator $Q^{\rm ext}_b$ acts on $L^2_b(m_r)$ as the
multiplication operator:
 \[
Q^{\rm ext}_b F(x(r))=x(r)F(x(r)).
 \]
It is known (see \cite{[15]}) that the multiplication operator is
a self-adjoint operator.

The Hilbert space $L^2_b(m_r)$ is the space of states of our
physical system in the coordinate representation. Since the
elements $e_n(y)\in {\cal H}$ are eigenfunctions of the
Hamiltonian $H=\frac 12 (aa^++a^+a)$, then $P_n(x(r))\in \hat
L^2_b(m_r)$ are eigenfunctions of the same Hamiltonian if its
action is considered on $L^2_b(m_r)$.

Recall that for different values of $b$ the sets \eqref{eq18} of
values of the coordinate are different. Moreover, for different
values of $b$ these sets do not intersect. Therefore, the spaces
$L^2_b(m_r)$ for them are different, since they consist of
functions defined on different sets.

\section{Eigenfunctions and spectra of the momentum operator}

By changing the basis $\{ e_n(y)\}$ by the basis $\{ e'_n(y)\}$,
where $e'_n(y)={\rm i}^{-n} e_n(y)$, we see that the momentum
operator $P={\rm i}(a^+-a)$ is given in the latter basis by the same
formula as the position operator is given in the basis $\{
e_n(y)\}$. This means that the operator $P$ is symmetric, but is
not self-adjoint. Moreover, it has infinitely many (in fact,
one-parameter family of) self-adjoint extensions.

Eigenfunctions of the momentum operator can be found (by using the
basis $\{ e'_n(y)\}$) in the same way as in the case of the
position operator. For this reason, we adduce only the results.

Eigenfunctions $\xi_p(y)$ of the momentum operator $P$,
 \[
P\xi_p(y)=p \xi_p(y),
 \]
are of the form
 \begin{gather*}
\xi_p(y) =\prod_{n=0}^\infty \left(1-2{\rm i}y p' {\breve
q}^{n+1}+y^2{\breve q}^{2n+2}\right) \\
\phantom{\xi_p(y)}{}
 =\left({\rm i}y\breve q
\left(\sqrt{1+{p'}^2}+p'\right);\breve q\right)_\infty \left(-{\rm i}y\breve q
\left(\sqrt{1+{p'}^2}-p'\right);\breve q\right)_\infty ,
 \end{gather*}
where $p':=\frac 12 (q-1)^{1/2}p$. The function $\xi_p(y)$ can be
decomposed in the $q ^{-1}$-Hermite polynomials $h_n(p|\breve q
)$:
 \begin{gather*}
\xi_p(y) =\sum_{n=0}^\infty \frac{{\rm i}^{-n}y^n{\breve
q}^{n(n+1)/2}}{(\breve q;\breve q)_n} h_n(p'|\breve q) \\
\phantom{\xi_p(y)}{}=\sum_{n=0}^\infty e_n(y) \frac{{\rm i}^n{\breve q}^{n(n+1)/4}}{(\breve
q;\breve q)_n^{1/2}}h_n(x'|\breve q).
 \end{gather*}
Thus, we have the following decomposition of the eigenfunctions
$\xi_p(y)$ in the basis elements \eqref{eq6} of the Hilbert space
${\cal H}$:
 \begin{gather}
\xi_p(y)=\sum_{n=0}^\infty \tilde P_n(x)e_n(y),  \label{eq21}
 \end{gather}
where the coefficients $\tilde P_n(x)$ are given by the formula
 \[
\tilde P_n(x)=\frac{{\rm i}^n{\breve q}^{n(n+1)/4}}{(\breve
q;\breve q)_n^{1/2}}h_n(x'|\breve q).
 \]

Using orthogonality relations for the polynomials $h_n(x'|\breve
q)$ described above, we can make the following conclusion:

 \begin{theorem}
Self-adjoint extensions $P^{\rm ext}_b$ of the position operator
$P$ are given by the parameter~$b$, $\breve q\le b<1$. The
spectrum of the extension $P^{\rm ext}_b$ coincides with the set
of points
 \begin{gather}
p_b(r)=(q^rb^{-1}-bq^{-r})/(q-1)^{1/2},\qquad r=0,\pm
1,\pm2,\ldots .       \label{eq22}
 \end{gather}
 \end{theorem}

This set coincides with the set of values of the momentum of our
physical system fixed by the parameter $b$. To the eigenvalues
\eqref{eq22} there are corresponding eigenfunctions
 \begin{gather}
\xi_{p_b(r)}(y)=\sum_{n=0}^\infty \tilde P_n(p_b(r))e_n(y),\qquad
r=0,\pm 1,\pm 2,\ldots.         \label{eq23}
 \end{gather}

It follows from \eqref{eq22} and from assertions of Section 2 that
the spectra of extensions of the operator $P$ have the same
properties as the spectra of extensions of the position operator
$Q$:

 \begin{corollary}
$({\rm a})$ Spectra of the self-adjoint extensions $P^{\rm ext}_b$
are discrete and simple.

$({\rm b})$ Spectra of two different self-adjoint extensions
$P^{\rm ext}_b$ and $P^{\rm ext}_{b'}$, $b\ne b'$, are mutually
separated.

$({\rm c})$ For a given real number $p_0$ there exists a (unique)
self-adjoint extension $P^{\rm ext}_b$ such that $p_0$ is a
spectral point of $P^{\rm ext}_b$.
 \end{corollary}

In the same way as in the case of the position operator we derive
that the eigenfunctions \eqref{eq23} constitute a basis of the
Hilbert space ${\cal H}$. Let us normalize it. To do this, we make
the same reasoning as in Section 7, and obtain that the functions
 \[
\xi^{\rm norm}_{p_b(r)}(y)  =m_r^{1/2}\xi_{p_b(r)}(y),\qquad
r=0,\pm 1,\pm 2,\ldots,
 \]
form a normalized basis of ${\cal H}$, where $m_r\equiv m_r(b)$ is
given by \eqref{eq15}.

\section[Momentum realization of ${\cal H}$]{Momentum realization of $\boldsymbol{{\cal H}}$}

In order to realize $P^{\rm ext}_b$ as a self-adjoint operator we
use the reasoning of Section 8, namely, we construct a one-to-one
isometry $\Omega'$ of the Hilbert space ${\cal H}$ onto the
Hilbert space $\hat L^2_b(m_r)$ of functions $F$ on the set of
points \eqref{eq22} (which coincides with the set of values of the
momentum) with the scalar product
 \[
\langle F(p_b(r)),F'(p_b(r))\rangle =\sum_{r=-\infty}^\infty
m_rF(p_b(r))\overline{F'(p_b(r))}.
 \]
It follows from \eqref{eq17} that the polynomials $\tilde
P_n(p_b(r))$ are orthogonal and constitute an orthonormal basis of
$\hat L^2_b(m_r)$. For a fixed $b$, the isometry $\Omega'$ is
given by the formula
 \[
\Omega': \ \ {\cal H}\ni f\to F(p_b(r))=m_r^{-1/2}\langle
f,\xi^{\rm norm}_{p_b(r)}(y) \rangle_{\cal H}\in \hat L^2_b(m_r).
 \]
It follows from \eqref{eq21} that
 \[
{\cal H}\ni e_n(y)\to m_r^{-1/2}\langle e_n(y),\xi^{\rm
norm}_{p_b(r)}(y) \rangle_{\cal H} =\tilde P_n(p_b(r)).
 \]
This formula shows that $\Omega'$ is indeed a one-to-one isometry.

The operator $P^{\rm ext}_b$ acts on $\hat L^2_b(m_r)$ as the
multiplication operator:
 \[
 P^{\rm ext}_b F(p_b(r))=p_b(r)F(p_b(r))
 \]
and this operator is self-adjoint.

The Hilbert space $\hat L^2_b(m_r)$ is the space of states of our
physical system in the momentum representation. Since the elements
$e_n(y)\in {\cal H}$ are eigenfunctions of the Hamiltonian
$H=\frac 12 (aa^++a^+a)$, then $\tilde P_n(p_b(r))\in \hat
L^2_b(m_r)$ are eigenfunctions of the same Hamiltonian if its
action is considered in $\hat L^2_b(m_r)$.

For different values of $b$ the sets \eqref{eq22} of values of the
momentum are different. Therefore, the spaces $\hat L^2_b(m_r)$
for them are different, since they consist of functions defined on
different sets. Clearly, we may identify $L^2_b(m_r)$ with $\hat
L^2_b(m_r)$.
 \medskip

{\bf Physical conclusion.} Our consideration shows that the
creation and annihilation operators~$a^+$ and $a$ of Section 3 at
$q>1$ cannot determine a physical system without further
indications. Namely, in order to determine a physical system we
have to take appropriate self-adjoint extensions of the operators
$Q$ and $P$. Thus, the $q$-oscillator algebra of Section 3 in fact
determine two-parameter family of $q$-oscillators. We denote them
by $O(b,b')$, $\breve q \le b,b'<1$, where $b$ and $b'$ are
determined by $Q^{\rm ext}_b$ and $P^{\rm ext}_{b'}$.

\section[Fourier transforms related to the $q$-oscillator with $q>1$]{Fourier transforms 
related to the $\boldsymbol{q}$-oscillator with $\boldsymbol{q>1}$}

Let us first consider what we have in the case of the usual
quantum harmonic oscillator. This oscillator is determined by the
relation
 \[
 AA^+-A^+A=1.
 \]
For position and momentum operators $Q_A$ and $P_A$ we have
 \[
 Q_A=A^+ +A,\qquad  P_A={\rm i}(A^+-A).
 \]
 The Hilbert space of states ${\cal H}_A$ is spanned by the
 orthonormal vectors
 \[
 |n\rangle,\qquad n=0,1,2,\ldots .
 \]
For eigenvectors of $Q_A$ and $P_A$ we have
 \[
 Q_A |x\rangle =p|x\rangle ,\qquad P_A |p\rangle =p|p\rangle
 \]
and ${\rm Spec}\ Q_A ={\mathbb R}$, ${\rm Spec}\ P_A ={\mathbb
R}$.

For $h\in {\cal H}_A$ we have
 \begin{gather}
\langle h,x\rangle =h(x),\qquad \langle h,p\rangle =\hat h(p).
 \label{eq22b}
 \end{gather}
In this way we obtain a realization of ${\cal H}_A$ as a space of
functions in the coordinate or as a space of functions in the
momentum. Then the functions $h(x)$ and $\hat h(p)$ from
\eqref{eq22b} are related with each other by the usual Fourier
transform:
 \[
h(x)=\frac{1}{\sqrt{2\pi}} \int^\infty_{-\infty} \hat h(p) e^{{\rm
i}px}dp.
 \]

An analog of this Fourier transform for the $q$-oscillator in
the case when $0<q<1$ is derived in \cite{[18]}. The aim of this
section is to give an analog of the Fourier transform for the
$q$-oscillator $O(b,b')$ for fixed $b$ and $b'$ (when $q>1$). This
analog is a transform on a discrete set since the coordinate and
the momentum run over discrete sets.

We fix $b$ and $b'$ from the interval $[ \breve q,1)$. Let $f\in
{\cal H}$ and
\begin{gather*}
\Omega f=F(x_b(r))\in L^2_b(m_r),
\\
 \Omega' f=\hat
F(p_{b'}(r'))\in \hat L^2_{b'}(m_{r'}).
\end{gather*}
We have to find a linear transform ${\cal F}: \hat
L^2_{b'}(m_{r'})\to L^2_{b}(m_{r})$ such that ${\cal F}\hat F= F$.
By the definition of $\Omega$ and $\Omega'$, one has
\begin{gather*}
F(x_b(r))=m_r^{-1/2}(b) \langle f, \varphi^{\rm
norm}_{x_b(r)}\rangle_{\cal H} ,
\\
 \hat F(p_{b'}(r'))=m_{r'}^{-1/2}(b') \langle f, \xi^{\rm
norm}_{p_{b'}(r')}\rangle_{\cal H} .
\end{gather*}
It is clear that
 \begin{gather*}
\varphi^{\rm norm}_{x_b(r)}(y) =\sum_{r'=-\infty}^\infty \langle
\xi^{\rm norm}_{p_{b'}(r')} , \varphi^{\rm
norm}_{x_b(r)}\rangle_{\cal H}
\xi^{\rm norm}_{p_{b'}(r')}(y) \\
\phantom{\varphi^{\rm norm}_{x_b(r)}(y)}{}= \sum_{r'=-\infty}^\infty {\cal
T}^{b'b}_{r'r} \xi^{\rm norm}_{p_{b'}(r')}(y) ,
 \end{gather*}
where ${\cal T}^{b'b}_{r'r}=\langle \xi^{\rm norm}_{p_{b'}(r')} ,
\varphi^{\rm norm}_{x_b(r)}\rangle$. Therefore,
 \begin{gather}
F(x_b(r)) =m_r^{-1/2}(b) \sum_{r'=-\infty}^\infty {\cal
T}^{b'b}_{r'r} \langle f, \xi^{\rm norm}_{p_{b'}(r')}\rangle \nonumber\\
\phantom{F(x_b(r))}{}=\sum_{r'=-\infty}^\infty
 \left( \frac{m_{r'}(b')}{m_r(b)}\right) ^{1/2}
 {\cal T}^{b'b}_{r'r} \hat F (p_{b'}(r')).
 \label{eq24}
 \end{gather}
Thus, an analog of the Fourier transform for the $q$-oscillator
$O(b,b')$ is given by the matrix $\left( {\cal F}^{b'b}_{r'r}
\right)_{r',r=-\infty}^{\infty}$. For entries of this matrix we
have
  \begin{gather*}
{\cal F}^{b'b}_{r'r} \equiv
\frac{m_{r'}(b')^{1/2}}{m_r(b)^{1/2}}{\cal T}^{b'b}_{r'r}=
\frac{m_{r'}(b')^{1/2}}{m_r(b)^{1/2}} \langle \xi^{\rm
norm}_{p_{b'}(r')} ,
\varphi^{\rm norm}_{x_b(r)}\rangle  \\
\phantom{{\cal F}^{b'b}_{r'r}}{}=m_{r'}(b') \langle
\xi_{p_{b'}(r')} , \varphi_{x_b(r)}\rangle\\ 
\phantom{{\cal F}^{b'b}_{r'r}}{}
=m_{r'}(b')
\sum_{n=0}^\infty \tilde P _n(p_{b'}(r'))P_n(x_b(r)) \\
\phantom{{\cal F}^{b'b}_{r'r}}{}=m_{r'}(b') \sum_{n=0}^\infty\frac{(-{\rm i})^n\breve q^{n(n+1)/2}}{
(\breve q;\breve q)_n}  h_n( \frac 12
(q^{r'}{b'}^{-1}-q^{-r'}b')|\breve q) h_n( \frac 12
(q^{r}{b}^{-1}-q^{-r}b)|\breve q)  .
 \end{gather*}
In order to sum up the last sum we set
 \[
q=e^\tau ,\qquad b=e^\sigma,\qquad b'=e^{\sigma'}.
 \]
Then
\begin{gather*}
\frac 12 (q^{r'}{b'}^{-1}-q^{-r'}b')=\sinh (\tau r'-\sigma'),
\\
\frac 12 (q^{r}{b}^{-1}-q^{-r}b)=\sinh (\tau r-\sigma).
\end{gather*}
Taking into account the relation
 \[
\sum_{n=0}^\infty h_n(\sinh \xi|\breve q)h_n(\sinh \eta|\breve q)
\frac{\breve q^{n(n-1)/2}}{(\breve q;\breve q)_n} R^n
=\frac{(-Re^{\xi+\eta},-Re^{-\xi-\eta},Re^{\xi-\eta},Re^{\eta-\xi};\breve
q)_\infty}{(R^2/\breve q ;\breve q)_\infty}
 \]
from \cite{[14]}, where $(a,b,c,d;\breve q )_\infty\equiv
(a;\breve q )_\infty (b;\breve q )_\infty (c;\breve q )_\infty
(d;\breve q )_\infty$, we derive that
 \begin{gather}
{\cal F}^{b'b}_{r'r}=\frac{m_{r'}(b')}{(-\breve q;\breve
q)_\infty}
 (\alpha^+_+;\breve q)_\infty (\alpha^-_-;\breve q)_\infty
 (\alpha^+_-;\breve q)_\infty (\alpha^-_+;\breve q)_\infty ,
  \label{eq25}
  \end{gather}
where
 \begin{gather*}
\alpha^+_+ ={\rm i} \breve q ^{r+r'+1}bb' , \\
\alpha^-_- ={\rm i} \breve q ^{-r-r'+1}b^{-1}{b'}^{-1} ,\\
 \alpha^+_- =-{\rm i} \breve q ^{r-r'+1}b{b'}^{-1} , \\
 \alpha^-_+ =-{\rm i} \breve q ^{-r+r'+1}b^{-1}b' .
 \end{gather*}

Thus, the Fourier transform ${\cal F}\equiv {\cal F}^{b'b}$,
corresponding to the $q$-oscillator $O(b,b')$, is given by formula
\eqref{eq24}, where entries of the matrix $\left( {\cal
F}^{b'b}_{r'r} \right)_{r',r=-\infty}^{\infty}$ are determined by
\eqref{eq25}. The inverse transform ${\cal F}^{-1}F=\hat F$ is
given by the formula
 \[
 \hat F (p_{b'}(r')) =\sum _{r=-\infty}^\infty
 \left( \frac{m_{r}(b)}{m_{r'}(b')}\right) ^{1/2}
\overline{{\cal T}^{b'b}_{r'r}}  F (p_{b}(r)),
 \]
where ${\cal T}^{b'b}_{r'r}$ are such as above. As in the case of
the usual Fourier transform, the corresponding Plancherel formula
holds.

\subsection*{Acknowledgements}

This research was partially supported by Grant 10.01/015 of the
State Foundation of Fundamental Research of Ukraine. Discussions
with N. Atakishiyev, I. Burban, and O. Gavrylyk are gratefully
acknowledged.

\LastPageEnding

\end{document}